\documentclass[twocolumn,floatfix,superscriptaddress,amsmath,showpacs,showkeys,aps,pre]{revtex4-1}

\usepackage[utf8]{inputenc}
\usepackage[final]{graphicx}
\usepackage{epsfig}
\usepackage{bm}
\usepackage[normalem]{ulem}
\usepackage[usenames]{color}
\usepackage{times}
\usepackage{verbatim}
\usepackage{xcolor}

\newcommand{\beq}{\begin{equation}}
\newcommand{\eeq}{\end{equation}}
\newcommand{\beqa}{\begin{eqnarray}}
\newcommand{\eeqa}{\end{eqnarray}}
\newcommand{\vecx}{\vec{x}}
\newcommand{\veck}{\vec{k}}
\newcommand{\vecp}{\vec{p}}
\newcommand{\ddk}{\frac{d^2 k}{(2\pi)^2}}
\newcommand{\ddp}{\frac{d^2 p}{(2\pi)^2}}

\begin{document}

\title{Glassy behavior of two-dimensional stripe-forming systems}

\author{Ana C. Ribeiro Teixeira}
\affiliation{Departamento de F\'{\i}sica,
Universidade Federal do Rio Grande do Sul, 
CP 15051, 91501-970 Porto Alegre, RS, Brazil}

\author{Daniel A. Stariolo}
\email{daniel.stariolo@ufrgs.br}
\affiliation{Departamento de F\'{\i}sica,
Universidade Federal do Rio Grande do Sul,
CP 15051, 91501-970 Porto Alegre, RS, Brazil}
\affiliation{National Institute of Science and Technology for Complex Systems}

\author{Daniel G. Barci}
\affiliation{Departamento de F{\'\i}sica Te\'orica,
Universidade do Estado do Rio de Janeiro, Rua S\~ao Francisco Xavier 524, 20550-013  
Rio de Janeiro, Brazil}

\date{\today}

\begin{abstract}
We study   two dimensional frustrated but non-disordered systems  applying  a replica approach to a stripe
forming model with competing interactions.   The phenomenology of the model is representative of several well known systems, like
high-Tc superconductors and ultrathin ferromagnetic films, which have been the subject of intense research.
We  establish the existence of a glass transition to a non-ergodic regime accompanied by an exponential
number of long lived metastable states, responsible for slow dynamics and non-equilibrium effects.
\end{abstract}



\maketitle

\section{Introduction}
\label{Introduction}
Many systems exhibit competition between different interactions. Competing interactions
 are frequently responsible for complex behavior, leading to slow dynamics,
metastability and energy landscapes characterized by a multiplicity of local minima, 
similar to spin and molecular glasses and other frustrated systems. Competing interactions
are also responsible for the appearance of complex patterns, like stripes, lamellae, bubbles and others~\cite{SeAn1995}.
Examples range from solid state systems, 
like ultrathin ferromagnetic films
~\cite{PoVaPe2006,AbVe2010} and strongly correlated electron liquids~\cite{KiFrEm1998,
FrKi1999}, to soft matter systems like Langmuir monolayers~\cite{SeMoGoWo1991},
block copolymers~\cite{HaChetal2002,RuBoBl2008} and
colloids~\cite{TaCo2006,ImRe2006,GlGrKaKoSaZi2007}. 

Although many characteristics of their phase diagrams and low temperature phases have been
widely investigated, there are still several important points which remain to be understood. 
Due to strong frustration effects it is very difficult to probe the equilibrium dynamics at low
temperatures, and usually long lived metastable states rule the physical behavior. This is particularly
dramatic in experiments, which often report effects of metastable phases and structures, although
this is not always properly recognized. A few experimental results quantifying
the low temperature dynamics of quasi-two-dimensional stripe forming systems have been reported\cite{HaChetal2002,AbVe2010}. 
Experimental and
also computer simulation results \cite{CaMiStTa2008,BoVi2002}  point to the presence of slow dynamics 
associated with the pinning of topological defects, which are the relevant excitations at low temperatures. 

There is a fascinating phenomenology in a family of  compounds that present high Tc superconductivity.
 In addition to superconductivity, typical ingredients found in systems with competing interactions, 
such as  inhomogeneity, anisotropy, disorder and glassiness, coexist. 
The deep understanding  of the interplay between all these complex phases is  a huge challenge form a  theoretical, 
as well as from an  experimental point of view\cite{KivFrad2003}.
The importance of  frustrated phase separation in  cuprates was early recognized \cite{EmKi1993}.  The intermediate state  
between the Mott insulator and the superconducting  phase is usually understood  as a spin glass  with local striped order, 
called ``cluster  glass''. 
Although the electronic cluster glass  state exhibits no known long range order, some electronic order is always detected 
by local probes \cite{cho1992,  pana2002,  curro2006}. 
Recently,  using atomic-resolution tunneling-asymmetry imaging, the cluster glass  was studied in detail for chemically different compounds\cite{Kohsaka2007, Lawler2010}. 
One of the main conclusions is  that the origin of this phase is the  intrinsic  electronic 
structure of the ${\rm CuO_2}$  planes, and not an extrinsic effect such as chemical doping or random impurities.   
In the same direction, recent measurements\cite{Daou2010} of the Nernst effect in $\rm YBa_2Cu_3O_y$, showed that the pseudo gap temperature coincides 
with the appearance of a strong in-plane anisotropy of electronic origin, compatible with the electronic nematic phase\cite{KiFrEm1998, Lawler2006}.  
Also, fluctuating stripes have been measured \cite{Parker2010} at the onset of the pseudogap state of  $\rm Bi_2Sr_2CaCu_2O_{8+x}$,  using 
spectroscopic mapping with a scanning tunneling microscope.  
Therefore, it seems that local inhomogeneity and/or anisotropy 
with slow dynamics is a rule in a wide sector of the  cuprates phase diagram.  

From the theoretical point of view, self-generated glassiness should be a relevant mechanism 
in any  stripe forming system, independently of the presence of quenched disorder\cite{ScWo2000}. 
 In Ref. \onlinecite{WeScWo2001}, by means of a recently developed replica method for dealing with 
frustrated systems without quenched disorder,   the existence of a glass transition
in three dimensional stripe forming systems was predicted.  
In this work we extend those calculations to a similar 
model in two spatial dimensions. There are several motivations to face this calculation. 
Firstly, the essential physics in the cuprates seems to be bi-dimensional and the same is true in ultrathin 
ferromagnets with perpendicular anisotropy, and many soft matter systems as cited above. On the other hand, there is a fundamental 
difference between  two and three dimensional models with competing interactions. While in  three dimensions the stripe order is quasi-long-ranged, 
due to logarithmically growing fluctuations ,  in the homogeneous two dimensional case there is no possible stripe order, since long distance fluctuations 
grow linearly.  We have found that this type of systems develop quasi-long-ranged orientational order\cite{BaSt2007,BaSt2009} with  nematic symmetry, 
a classical version of the same phase observed in the pseudo gap regime of cuprates.  Thus, the equilibrium state in two and three dimensional frustrated models 
are quite different. Therefore, a natural question is about the dynamics to approach equilibrium in these systems.  
To our knowledge, from a technical point of view, this is the first application
of the replica technique to a  uniformly frustrated  two-dimensional stripe forming model.
Our results show, despite the ordered phase is only locally striped  or  nematic, the  existence of a dynamical transition
to a non-ergodic regime at low temperatures. 
The calculation presented here complements other efforts to get a complete phase diagram
of two dimensional systems with competing interactions at different scales~\cite{BaSt2009,BaSt2011}.

In Section II we introduce the model and the essential technical background for the calculations, i.e. the replica approach
to uniformly frustrated systems together with the self-consistent screening approximation (SCSA). The details of the
replica technique for uniformly frustrated systems and the SCSA are well 
documented in the literature, so we decided not to give a detailed derivation of it. Instead we include a Supplementary
Material where the main steps leading to the SCSA in replica space are summarized. 
In Section III we show the main results related with the existence of a dynamical transition
to a glassy state. In Section IV we make a discussion of the present status of the phase diagram of
these models and close with some conclusions.

\section{Model and methods}
\label{model}

Usually,  competing interactions at different scales lead to the appearance of a momentum scale $k_0$ that 
dominates the low energy physics.  The simplest two-dimensional effective Hamiltonian that it is possible 
to write down with this behavior 
can be split into a quadratic  and an interaction part , ${\cal H}={\cal H}_0+ {\cal H}_i$.
The quadratic or ``free'' component can be written (in momentum space) as ~\cite{BaSt2007}:
\beq \label{model1}
{\cal H}_0=\int \frac{d^2k}{(2\pi)^2} \phi(\veck) \left( r_0+J(k^2-k_0^2)^2\right) \phi(-\veck) 
\eeq
where $r_0(T) \propto (T-T^*)$ and $T^*$ is the mean field critical temperature of the model,
 $\phi(\veck)$ is a scalar field, and the scale $k_0$ comes from the competition between interaction
terms at different scales~\cite{SeAn1995}. 
Note that we are considering systems with nearly isotropic interactions and then the kernel
depends on $k=|\veck|$.
The simplest interaction term is given by a local quartic term of the form
\beq \label{model2}
{\cal H}_i=v \int \left(\prod_{i=1}^4 \frac{d^2k_i}{(2\pi)^2}\right) \phi(\veck_1)\ldots\phi(\veck_4) 
\delta^2(\veck_1+\veck_2+\veck_3+\veck_4) 
\eeq
where $v$ measures the interaction intensity. The correlation function of the free part is given by:
\beq
{\cal G}_0^{-1}(k)  =  r_0+J(k^2-k_0^2)^2. 
\eeq
The free correlation is renormalized by the interaction term (\ref{model2}). The simplest correction
is given by the self-consistent field approximation in which the quartic term is approximated in the form
$\phi^4(\vec x)\simeq \langle \phi^2(\vec x)\rangle \phi^2(\vec x)$. In this way
the original effective theory is approximated by one which is quadratic in the fields $\phi$ and can be solved 
exactly. This amounts to renormalize the temperature dependence of the parameter $r_0$ giving~\cite{chaikin-1995}:
\beq
r(T) = r_0(T) +vT \int \ddk {\cal G}_0(\veck),
\eeq
where the (renormalized) free correlation function in the disordered phase is:
\beqa \label{freecorr}
{\cal G}_0^{-1}(k) & = & r+J(k^2-k_0^2)^2 \\
              &=& J(k^2-\alpha^2)(k^2-(\alpha^*)^2) \label{prop}
\eeqa
with roots given by $\alpha=k_0\sqrt{1+\frac{i}{k_0^2}\sqrt{\frac{r}{J}}}$.

The free correlation (\ref{freecorr}) can be Fourier transformed exactly, yielding in two dimensions:
\beq \label{barecorr}
G_0(x)=\frac{1}{8\pi i J\alpha_R\alpha_I}\left\{K_0(-i\alpha x)-K_0(i\alpha^*x)
             \right\}
\eeq
where $x=|\vec x|$, $\alpha_R$ and $\alpha_I$ are respectively the real and imaginary 
parts of $\alpha$ and $K_0(z)$ is a Hankel function. Asymptotically, for large $x$,
it behaves as:
\beqa
G_0(x \gg 1)&=&\frac{1}{\sqrt{32\pi}J\alpha_R\alpha_I (\alpha_R^2+\alpha_I^2)^{1/4}}
\times \nonumber \\
&\times&   \frac{e^{-\alpha_I x}}{x^{1/2}}
                  \sin{\left[\alpha_Rx-\frac{1}{2}arg(x)+\frac{\pi}{4}\right]}\; .
\eeqa
From this expression one easily identifies $\alpha_I=1/\xi$ as the inverse of the correlation
length and $\alpha_R=k_m$ as a modulation wave vector. These are the two natural 
characteristic scales in the high temperature phase of the system. $\alpha_R$ is weakly
dependent on temperature. To leading order in the small parameter $r/Jk_0^4$ one finds:
\beq
\alpha \approx k_0\left( 1+i\frac{1}{2k_0^2}\sqrt{\frac{r}{J}}\right)
\eeq
We see that, to leading order, the modulation wave vector is constant $\alpha_R \approx k_0$
and the correlation length is large, $\alpha_I \approx (1/2k_0)\sqrt{r/J} \ll 1$.

\subsection{Replica technique for uniformly frustrated systems}
\label{replicas}
As discussed in the introduction, systems with competing interactions at different scales, like
low dimensional electronic liquids and ultra-thin ferromagnetic films with strong perpendicular
anisotropy, have many metastable states
as a consequence of frustration induced by competition of interactions. There are many reports
in the literature showing metastable patterns and slow dynamics at low temperatures, which may
be related with glassy physics~\cite{HaChetal2002,AbVe2010,CaMiStTa2008,BoVi2002,Kohsaka2007, Lawler2010}. Then it is important to assess the relevance of metastable states
for the behavior of thermodynamic and dynamic functions. One way to do this is to compute the
possible existence of persistent long time correlations by means of a replica approach specially
devised to deal with systems without quenched disorder, but which nevertheless show signatures
of glassy physics, like the ones we are interested in. This technique, introduced and developed in
Refs.  \onlinecite{Monasson1995,MePa1999}, has been applied to a three dimensional Coulomb frustrated system
in Refs. \onlinecite{ScWo2000,WeScWo2001}. The essential idea of the method is to introduce a kind of ``pinning field''
$\psi(\vec x)$, which selects the local metastable (disordered) configurations by enhancing the weight of these
configurations in the partition function:
\begin{eqnarray}
Z[\psi,\beta]&=&g^{-1/2}\int {\cal D}\phi\times \nonumber \\
&\times& \exp\left\{-\frac{{\cal H}[\phi]}{T}-
                      \frac{g}{2}\int d^dx\,[\psi(\vec x)-\phi(\vec x)]^2\right\}
\end{eqnarray}
where the coupling $g\to 0^+$ should be taken after the thermodynamic limit. In order to take into
account the (possibly) many metastable configurations, one has to scan for all the configurations of 
the field $\psi$. This can be done by introducing replicas, which leads to a replicated free energy:
\beq \label{freplica}
F_{\psi}(m,\beta)=\lim_{g\to 0^+}-\frac{1}{\beta m}\ln{\left(\int d\psi \ Z^m[\psi,\beta]\right)}
\eeq
In the end, the limit $m \to 1$ must be taken. Details of the method have been extensively 
described in the literature \cite{Monasson1995,MePa1999,WeScWo2001}, so we refer the reader interested in the details
of the method to consult those references. The essential point here is that, if the system has
an exponentially large number of metastable states in the thermodynamic limit, then the replicated
free energy will show the usual contribution plus a new one, of entropic nature, which allows to define a
{\em configurational entropy} as:
\beq \label{sconf}
S_c(\beta)=\beta \left[f-F_{\psi}(m=1,\beta)\right],
\eeq
where $f$ is the equilibrium free energy of the system.
A finite configurational entropy is then associated with glassy behavior, which can be inferred from the
long-time behavior of dynamical correlation functions. The correlation functions in
the replicated theory obey a Dyson equation:
\beq \label{dyson}
{\cal G}_{ab}^{-1}(\vec k)={\cal G}_0^{-1}(\veck)\delta_{ab}+\Sigma_{ab}(\vec k)-\frac{g}{\beta m},
\eeq
where $a,b$ are replica indices.
Here, glassy physics is associated with the existence of finite off-diagonal elements in the replica
self-energy matrix $\Sigma_{ab}$. Following Westfahl et al.\cite{WeScWo2001}, we use for ${\cal G}_{ab}(\veck)$ the following
ansatz:
\beq \label{gansatz}
{\cal G}_{ab}(\veck)=[{\cal G}(\veck)-{\cal F}(\veck)]\delta_{ab}+{\cal F}(\veck), 
\eeq
in such a way that  ${\cal F}(\veck)$ parametrizes the off-diagonal elements of  ${\cal G}_{ab}(\veck)$.  
The diagonal elements in replica space correspond to static correlations ${\cal G}(\vecx-\vecx')=
\beta\langle \phi(\vecx)\phi(\vecx')\rangle$. The meaning of the off-diagonal elements in replica space can
be elucidated comparing the long time stationary solution of the Langevin dynamics of the system with its
canonical equilibrium properties~\cite{WeScWo2001,Crisanti2008}. Off-diagonal elements correspond to the long time
limit of dynamical correlations ${\cal F}(\vecx-\vecx')=\lim_{t \to \infty} \beta \langle \phi(\vecx,t)\phi(\vecx',0)\rangle$. 
It can be shown that in a disordered system with thermodynamic equilibrium solution with one step of
replica symmetry breaking (1RSB), the same theory leads to the long time limit of dynamical correlations
by taking the limit of the replica parameter $m \to 1$~\cite{Crisanti2008}. Inserting (\ref{gansatz}) into (\ref{dyson})
one gets for the diagonal elements in the limit $m \to 1$:
\beq
{\cal G}^{-1}(\veck)={\cal G}_0^{-1}(\veck)+\Sigma_{\cal G}(\veck)
\label{gcorr}
\eeq
and for the off-diagonal elements:
\beq \label{fcorr}
{\cal F}(\veck)={\cal G}(\veck)-\frac{{\cal G}(\veck)}{1-\Sigma_{\cal F}(\veck){\cal G}(\veck)}\equiv {\cal G}(\veck)-{\cal K}(\veck)
\eeq
in which a new function ${\cal K}(\veck)$ has been defined which measures the depart from liquid or disordered 
behavior. In Eqs. (\ref{gcorr}) and (\ref{fcorr}), $\Sigma_{\cal G}$ and $\Sigma_{\cal F}$ are the diagonal and off-diagonal 
self-energies respectively. 

At this point it is important to note that a linear, perturbative approach for the self-energy matrix
$\Sigma_{ab}$ is unable to give any glassy physics, leading to zero off-diagonal elements in the limit $g \to 0$. 
Then, it is necessary to go beyond the linear (Hartree) approximation in order to test for a possible glassy phase.
It turns out that a self-consistent screening approximation (SCSA), which amounts to sum an infinite class
of diagrams exactly, can do the job (see \onlinecite{WeScWo2001} and Supplementary Material). In the following we briefly describe the steps
for computing the off-diagonal elements of the correlation matrix within the SCSA for the system described
by static correlations given by (\ref{barecorr}).

\subsection{The self-consistent screening approximation}
\label{scsa}

The set of self-consistent equations for the two point correlation function
from the replica approach in the SCSA is given by (see Supplementary Material):
\beqa
{\cal G}_{ab}^{-1}(\vec k)&=&\left({\cal G}_0^{-1}(\veck)+\Sigma_{{\cal G}}(\vec k)\right)
        \delta_{ab}+\Sigma_{{\cal F}}(\vec k)(1-\delta_{ab}), \\
({\cal G}_0^{-1}(\veck))_{ab}&=&\left[r+J(k^2-k_0^2)^2\right]\delta_{ab}, \\
\Sigma_{ab}(\veck)&=&\frac{1}{2}\int \ddp {\cal D}_{ab}(\vec p){\cal G}_{ab}(\vec p
         +\veck), \label{fselfe}\\
{\cal D}_{ab}(\veck)&=&\left. \frac{v}{1+v\Pi(\veck)}\right|_{ab},\\
\Pi_{ab}(\veck)&=&\int \ddp {\cal G}_{ab}(\vecp){\cal G}_{ba}(\vecp+\veck).
\eeqa
 The aim of the calculation is to compute the off-diagonal self-energy $\Sigma_{{\cal F}}(\vec k)$. A non
zero value of this function in some temperature interval then signals the presence of a regime with
ergodicity breaking and glass-like characteristics. As a by-product of the calcultation, if an ergodic-non
-ergodic transition is found, the configurational entropy can be computed, which gives information on the
multiplicity of metastatble states in the free energy of the system.

The diagonal part of the polarization function is
\beq
\Pi_{\cal G}(\veck)=\int \ddp {\cal G}(\vecp){\cal G}(\vecp+\veck).
\eeq
 With ${\cal G}(\veck)$ given by (\ref{prop}) the diagonal polarization can be calculated exactly, giving
\beqa
\Pi_{\cal G}(k)&=&\frac{1}{\pi J^2[\alpha^2-(\alpha^*)^2]}  \times \nonumber \\
&\times&
          \left[\frac{\sinh^{-1}{\left(\frac{ik}{2\alpha}\right)}}{k\sqrt{k^2-4\alpha^2}}+
\frac{\sinh^{-1}{\left(\frac{-ik}{2\alpha^*}\right)}}{k\sqrt{k^2-4(\alpha^*)^2}}-\right. \nonumber \\
   & &2\left. \frac{\sinh^{-1}{\left[\frac{1}{2|\alpha|}\sqrt{k^2-(\alpha+\alpha^*)^2} \right]}}{\sqrt{[k^2-(\alpha+\alpha^*)^2][k^2-(\alpha-\alpha^*)^2]}} \right]
\eeqa
Expanding in real and imaginary parts to leading order in $\alpha_I$, one arrives 
(for $x=k/2k_0 < 1$) at the simple expression:
\beq \label{pig}
\Pi_{\cal G}(k)=\frac{1}{16\pi J^2k_0^4\alpha_I^2}\frac{\sin^{-1}{\sqrt{1-x^2}}}{x\sqrt{1-x^2}}.
\eeq
The off-diagonal polarization is defined as
\beq
\Pi_{\cal F}(\veck)=\int \ddp {\cal F}(\vecp){\cal F}(\vecp+\veck)
\eeq
As discussed in Ref.  \onlinecite{WeScWo2001},  $\Sigma_{{\cal F}}$ is weakly dependent on
wave vector.  Then, from the form of
equations (\ref{fcorr}) it follows that the function ${\cal K}(\veck)$ has essentially the same functional
form of ${\cal G}(\veck)$, i.e. ${\cal K}^{-1}(\veck)=z+J(k^2-k^2_0)^2$ with 
a different correlation length. This implies that
\beq \label{sigmafans}
\Sigma_{{\cal F}}=r-z.
\eeq
With these definitions we find at leading order:
\beq \label{pif}
\Pi_{\cal F}(k)=\frac{1}{16\pi J^2k_0^4}\left(\frac{1}{\alpha_I}-\frac{1}{\beta_I}\right)^2
                        \frac{\sin^{-1}{\sqrt{1-x^2}}}{x\sqrt{1-x^2}},
\eeq
where $\beta_I = (1/2k_0)\sqrt{z/J}$ is the inverse correlation length of the function ${\cal K}$.
In the polarization functions (\ref{pig}) and (\ref{pif}) the variable $x$ is limited from below
by a small cutoff of order $\epsilon=\alpha_I-\beta_I$. Also note that the function 
$\frac{\sin^{-1}{\sqrt{1-x^2}}}{\sqrt{1-x^2}}$ is always of order one for $0\leq x\leq 1$.
With these reasonable simplifications, the expressions for the diagonal and off-diagonal
polarization functions get the same form as those found in the three dimensional model\cite{WeScWo2001}. 
This is a consequence of the form in which the bare correlation
(\ref{prop}) depends on wave vector $\veck$, {\em i.\ e.\ },  for $\alpha_I,\beta_I\ll1$ the integrals are dominated by the scale
$k_0$ and then the dimension-full integration measure  only modifies a constant pre-factor, but not the $\alpha_I,\beta_I$ dependence.    
Then, approximating the off-diagonal self-energy by its value at the modulation wave vector $k_0$, 
equation (\ref{fselfe}) gives:
\beq \label{sigmaf}
\Sigma_{\cal F}(k_0)\approx -\pi Jk_0^3\alpha_I^2 \frac{\left(1-\frac{\alpha_I}{\beta_I}\right)^2}
                        {1-\left(1-\frac{\alpha_I}{\beta_I}\right)^2}\left(\frac{1}{\alpha_I}-
               \frac{1}{\beta_I}\right)
\eeq
Together with (\ref{sigmafans})
this allows to close the self-consistent equations for the off-diagonal self-energy function,
a relation which encodes a possible ergodicity breaking transition.

\section{Results}
\label{results}

\subsection{Ergodicity breaking transition}
\begin{figure}[!htb]
\begin{center}
\includegraphics[scale=0.35,angle=-90]{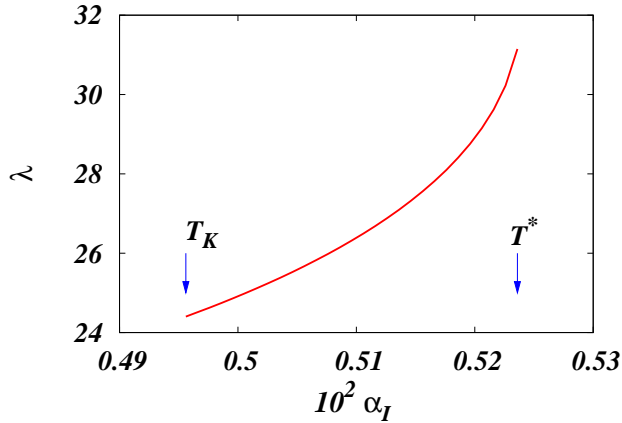}
\caption{(Color online) The wandering length as a function of the inverse correlation length
for $k_0=0.1$, $J=1/k_0^2$ and $v=0.001$.}
\label{lambda}
\end{center}
\end{figure}

From (\ref{sigmafans}) and remembering that $\alpha_I = (1/2k_0)\sqrt{r/J}$ and a similar
expression for $\beta_I = (1/2k_0)\sqrt{z/J}$, we can rewrite $\Sigma_{{\cal F}}=4Jk_0^2(\alpha_I^2-\beta_I^2)$,
which combined with expression (\ref{sigmaf}) gives the following algebraic equation in the parameters
$\alpha_I$ and $\beta_I$:
\beq
\beta_I^2-\alpha_I^2=\frac{\pi k_0 \alpha_I^2}{4}\frac{\left(1-\frac{\alpha_I}{\beta_I}\right)^2}
                        {1-\left(1-\frac{\alpha_I}{\beta_I}\right)^2}\left(\frac{1}{\alpha_I}-
               \frac{1}{\beta_I}\right)
\eeq
Factorizing the trivial (liquid) solution $\beta_I^*=\alpha_I^*$ and defining $\delta=\frac{4\alpha_I}{\pi k_0}\ll 1$ 
one finds another, non trivial solution $\beta_I^* \approx 3\alpha_I^*$. This solution implies the existence
of a transition to a non ergodic regime, in the sense that the dynamic correlations ${\cal F}$ have
a persistent part in the long time limit. This is the main result of this work. 
At the transition temperature, the correlation length is
\beq
\xi^*=1/\alpha_I^*=60/\pi k_0
\eeq
 With the modulation length given by $l_m=2\pi /k_0$, 
the ratio between the
correlation and modulation lengths at the transition is approximately $\xi^*/l_m\approx 3$. One can also
define a third characteristic length, associated with the onset of long time correlations. Noting that the
off-diagonal self-energy has dimensions of wave vector squared, it is possible to introduce  
a {\em wandering length} $\lambda$ defined by\cite{WeScWo2001}:
\beq
\Sigma_{{\cal F}}=-\lambda^{-2}=4Jk_0^2(\alpha_I^2-\beta_I^2).
\eeq
This new quantity can be interpreted as the length scale up to which topological defects of the
stripe structure can move. It is expected that 
in the high temperature phase the system is in a fluid-like phase and defects can wander without limit. In this regime
 $\beta_I=\alpha_I$ and then $\lambda=\infty$. If a transition to a non ergodic regime happens at some
temperature, then $\lambda$ will be finite. As shown in figure \ref{lambda}, $\lambda$ increases with temperature
and attains a finite value at the ergodic-non-ergodic transition, where it jumps to infinity in the fluid phase. 
In our model, we find at the transition a value $\lambda_{glass}\approx 10/(\pi J^{1/2}k_0^2)$. 
\subsection{Configurational entropy}
\begin{figure}[!htb]
\begin{center}
\includegraphics[scale=0.35,angle=-90]{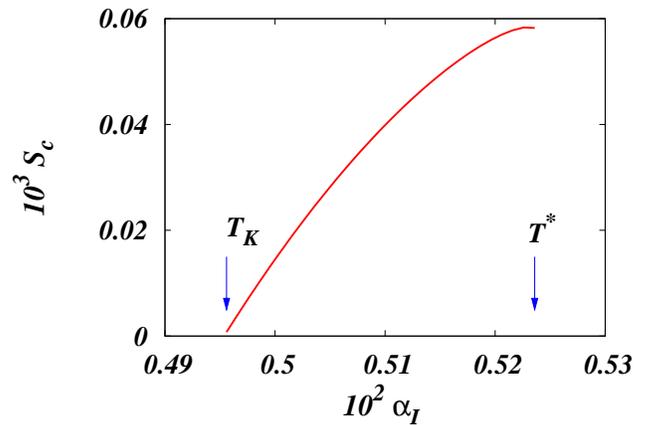}
\caption{(Color online) Configurational entropy as a function of the inverse correlation length. 
The parameters were fixed to $k_0=0.1$,
$J=1/k_0^2$ and $v=0.001$.
}
\label{entropy}
\end{center}
\end{figure}

Provided the existence of a transition to a non-ergodic regime has been established, we address the
calculation of the number of metastable configurations, or configurational entropy.
From (\ref{freplica}) and (\ref{sconf}) the configurational entropy can be obtained as
\beq
S_c=\beta \left. \frac{\partial F_{\psi}(m)}{\partial m}\right|_{m=1}.
\eeq
Within the SCSA the replicated free energy is given by
\beq
2m\beta F_{\psi}=Tr \ln{{\cal G}^{-1}}+Tr \ln{{\cal D}^{-1}}-Tr (\Sigma {\cal G}).
\eeq
Deriving with respect to $m$ and taking $m=1$ at the end, $S_c$
can be written as $S_c=s_c^{(1)}+s_c^{(2)}$ with:
\beq
s_c^{(1)}=-\frac{1}{2}\int \frac{d^2k}{(2\pi)^2}\left\{\ln{\left(1-\frac{{\cal F}(\veck)}{{\cal G}(\veck)}\right)}+
                           \frac{{\cal F}(\veck)}{{\cal G}(\veck)} \right\}
\eeq
and
\beqa
s_c^{(2)}&=&\frac{1}{2}\int \frac{d^2k}{(2\pi)^2}\times \nonumber \\
&\times& \left\{\ln{\left(1-\frac{v\Pi_{{\cal F}}(\veck)}{1+v\Pi_{{\cal G}}(\veck)}\right)}+
                           \frac{v\Pi_{{\cal F}}(\veck)}{1+v\Pi_{{\cal G}}(\veck)} \right\}.
\eeqa
Performing the integrals we obtain
\beq
s_c^{(1)}=\frac{k_0\beta_I}{4}\left(1-\frac{\alpha_I}{\beta_I}\right)^2
\eeq
and
\beqa
s_c^{(2)}&=&\frac{1}{(16\pi J^2k_0^4)^2}\frac{v^2k_0^2}{2\pi \alpha_I^4} \times \nonumber \\
&\times&              
 \left\{ \left(1-\frac{\alpha_I}{\beta_I}\right)^2+\ln{\left[1-\left(1-\frac{\alpha_I}{\beta_I}\right)^2\right]}\right\}.
\eeqa
The first term is always positive but the second can be negative as the temperature decreases from the 
transition point. At some point the second negative term dominates over the first term and the configurational
entropy becomes negative. As the entropy cannot be negative, this second characteristic temperature has been interpreted
as signalling a transition
to an ideal glass state. Below this temperature, called Kauzmann temperature ($T_K$) in the glass transition literature, 
the configurational entropy is
zero and the system freezes into an amorphous state, which is, from this temperature down, a thermodynamically stable state. Then there is a temperature window, between $T_K$ and $T_{glass}$,
in which the system has an exponentially large number of metastable states, as reflected in a finite value of $S_c$.
In figure \ref{entropy} we show the behavior of $S_c$ with the
inverse correlation length $\alpha_I$. As in figure \ref{lambda} the numerical values of the constants were fixed to $k_0=0.1$,
$J=1/k_0^2$ and $v=0.001$.

\section{Discussion and conclusions}
\label{conclusions}
We have applied a replica approach for frustrated but non-disordered systems to a two dimensional stripe
forming model with competing interactions. The model is representative of several well known systems, like
high-Tc superconductors and ultrathin ferromagnetic films, which have been the subject of intense research.
We have established the existence of a glass transition to a non-ergodic regime accompanied by an exponential
number of long lived metastable states, responsible for slow dynamics and non-equilibrium effects. From a
technical point of view, these
results may come as no surprise since they are very similar to those already known in the three dimensional 
version of the model. Nevertheless,
as stated in \S\ref{Introduction}, the possible existence of glassy behavior in two dimensions is specially
interesting due to the very different nature of the equilibrium phases of the systems in $d=2$ and $d=3$. 
From an equilibrium point of view, the existence of a continuous symmetry in $d=2$ associated with the isotropic
nature of interactions implies that no long range order can survive. Instead of the stripe phase found in
three dimensions, the relevant phase in two dimensions is a nematic one\cite{BaSt2007,BaSt2009} ,
with quasi-long-range orientational order and only short range translational order down to zero temperature.
The nematic phase in a stripe forming system is characterized by a proliferation of topological defects
which can naturally lead to an arrest of the dynamics as the temperature is lowered. In Refs. \onlinecite{BaSt2007,BaSt2009}
we showed that symmetry considerations allow the existence of other interaction terms in the Landau-Ginzburg
expansion of the model defined in (\ref{model1}) and (\ref{model2}), leading to possible anisotropic 
phases of nematic character. The temperature at which the isotropic-nematic transition takes place was
found to be proportional to   the nematic order parameter couplings in the
Landau expansion. These terms, although essential for the equilibrium behavior of the system, are not
relevant for searching a possible freezing of the isotropic, high temperature correlations, as pursued in
the present work.  Thus, there are at least  two characteristic  temperatures determining, on one hand, the equilibrium isotropic/nematic 
phase transition and, on the other hand, a glassy non-equilibrium behavior of the system. 
A natural question now is to know the relation between both temperatures,
which implies determining the dependence of the phenomenological nematic coupling constant  on more
microscopic parameters of the model, like $J$  and $k_0$. Some work in this direction has already been done
\cite{BaSt2011} but the question is still not completely settled. Having a more microscopic model to begin
with would allow to make quantitative predictions which could be in principle contrasted with experiments.
Nevertheless the very existence and relevance of the nematic phase is still controversial and the existence
of a glass-like transition adds a new ingredient to the theoretical understanding of low dimensional stripe
forming systems.

\acknowledgments
The Brazilian agencies, {\em Funda\c c\~ao de Amparo \`a Pesquisa do Estado do Rio
de Janeiro}  (FAPERJ) and {\em Conselho Nacional de Desenvolvimento Cient\'\i
fico e Tecnol\'ogico} (CNPq) are acknowledged  for partial financial support.

%

\end{document}


\begin{center}
{\large \bf Supplementary information}

\vspace{0.5cm}

{\large \bf Glassy behavior of two-dimensional stripe-forming systems}

\vspace{0.5cm}

{\bf Ana C. Ribeiro-Teixeira, Daniel A. Stariolo and Daniel G. Barci}
\end{center}

\vspace{0.8cm}

Here we include some details about the Self-consistent Screening Approximation within which we obtain the configurational entropy.
\vspace{0.5cm}

The SCSA is an improvement on the Hartree self-consistent field approximation\cite{Chaikin}. It may be easily shown that the Hartree mean-field approximation for $\langle \phi(x)^2 \rangle$ becomes exact in the spherical limit, when the limit $n\rightarrow\infty$ is taken for the number of components of the (vectorial) field $\phi_i(x)$ in $O(n)$ models. Extensivity of the Ginzburg-Landau functional Hamiltonian requires that the coupling be $v=\mathcal{O}(1/n)$. The first perturbative correction to the Hartree propagator is then $\mathcal{O}(1/n)$, which is the SCSA contribution.

Resummation of the diagrams up to $\mathcal{O}(1/n)$ in Dyson's equation for the propagator $\mathcal{G}_{ab}(k)$ gives the three diagrams shown in Figure 2, where the wiggly line corresponds to the dressed interaction $\mathcal{D}(k)$, containing all possible $\mathcal{O}(1/n)$ contributions (see Figure 1).

\begin{center}
\begin{picture}(50,30)(0,-3)
\Photon(25,0)(50,0){1}{8}
\label{fig_D}
\end{picture} $\,\,\,=\!\!\!\!$ \begin{picture}(50,30)(0,-3)
\DashLine(25,0)(50,0){3}
\end{picture} $\,\,\,+\!\!$ \begin{picture}(55,30)(0,-3)
\DashLine(12.5,0)(25,0){3}
\GOval(31.25,0)(3.5,6.25)(0){1}
\Photon(37.5,0)(55,0){1}{6}
\end{picture}
\Text(12,-15)[]{{\bf Figure 1}}
\end{center}
\vspace{0.5cm}

\begin{center}
\begin{minipage}[t]{0.32\linewidth}
\begin{picture}(85,80)(0,0)
\Line(25,30)(85,30)
\PhotonArc(55,30)(10,0,180){1}{7}
\Text(55,20)[]{(a)}
\end{picture}
\end{minipage}
\begin{minipage}[t]{0.32\linewidth}
\begin{picture}(95,80)(0,0)
\Line(25,30)(95,30)
\DashLine(60,30)(60,50){3}
\CArc(60,61)(11,180,0)
\CArc(60,61)(11,0,180)
\Photon(49,61)(71,61){1}{6}
\Text(60,20)[]{(b)}
\end{picture}
\end{minipage}
\begin{minipage}[t]{0.32\linewidth}
\begin{picture}(95,80)(0,0)
\Line(25,30)(95,30)
\DashLine(60,30)(60,50){3}
\CArc(60,61)(11,180,0)
\CArc(60,61)(11,0,180)
\Text(60,20)[]{(c)}
\Text(55,0)[]{{\bf Figure 2}}
\end{picture}
\end{minipage}
\end{center}

\begin{center}
$r-r_0=\,\,$\begin{picture}(30,50)(0,-3)
\DashLine(15,0)(15,15){3}
\GCirc(15,25){10}{1}
\end{picture} $\,\,\,+$ \begin{picture}(30,50)(0,-3)
\DashLine(15,0)(15,15){3}
\GCirc(15,25){10}{1}
\CArc(15,25)(11,270,360)
\Photon(5,25)(25,25){1}{8}
\Text(100,0)[]{(a)}
\end{picture}$=v\,\mbox{Tr}_{\mathbf{q}}\,\mathcal{G}(\mathbf{q})$
\vspace{0.75cm}

\begin{picture}(30,30)(0,-3)
\Line(0,0)(30,0)
\Line(0,1.5)(30,1.5)
\end{picture} $\,\,\,=\,\,$ \begin{picture}(30,20)(0,-3)
\Line(0,0)(30,0)
\end{picture} $\,+\,$ \begin{picture}(60,30)(0,-3)
\Line(0,0)(40,0)
\PhotonArc(30,0)(10,0,180){1}{7}
\Line(40,0)(60,0) 
\Line(40,1.5)(60,1.5)
\Text(85,2)[]{(b)}
\Text(70,-20)[]{{\bf Figure 3}}
\end{picture}
\end{center}
\vspace{1cm}

The SCSA introduces a non-local contribution (diagram 2-a) which will result in a non-trivial configurational entropy. One recognizes from diagrams 2-b and c the Hartree (local) propagator. The self-energy is then given by the irreducible part of diagram 2-a. It is worth noting that the Hartree mass term is also self-consistently renormalized within SCSA (see Figure 3-a).

For the study of the glassy phase, we have to give some non-trivial matrix structure to the propagator in replica space, reminiscent of the dynamic nature of this phase. The simplest non-trivial structure is the replica symmetric (RS) one: one associates a (static) correlation to the diagonal elements, and a (long-time) dynamic correlation to the off-diagonal ones\cite{WeScWo2001,Monasson1995}. The configurational entropy may then be associated to the appearance of a non-trivial matrix structure in replica space.

Together with the Dyson's equation (\ref{propag}) below and Figure 3-b (above), the functions entering the approximation are given by:
\begin{eqnarray}
&&\mathcal{G}_{ab}^{-1}(\mathbf{k})=(\mathcal{G}_0(k))_{ab}^{-1}+\Sigma_{ab}(\mathbf{k})
\label{propag}\\
&&(\mathcal{G}_0(\mathbf{k}))_{ab}^{-1}=\left[r+J(k^2-k_0^2)^2\right]\delta_{ab}
\nonumber\\
&&\Sigma_{ab}(\mathbf{k})= \int \frac{d^2 p}{(2\pi)^{2}}\, \mathcal{D}_{ab}\, (\mathbf{p}) \,\mathcal{G}_{ab}(\mathbf{p}+\mathbf{k})\nonumber\\
&&\mathcal{D}_{ab}(\mathbf{k})=v\left[1+v\Pi(\mathbf{k})\right]_{ab}^{-1}
\label{dressed}\\
&&\Pi_{ab}(\mathbf{k})= \int \frac{d^2 p}{(2\pi)^2}\, \mathcal{G}_{ab}(p)\, \mathcal{G}_{ba}(\mathbf{p}+\mathbf{k}),\nonumber
\end{eqnarray}
\noindent{}where $\mathcal{G}_0(\mathbf{k})$ is the Hartree propagator. The mass term $r=r_0+\,\, v\, \mbox{Tr}_{\mathbf{q}}\,\mathcal{G}(\mathbf{q})$ (see Figure 3-a) entering $\mathcal{G}_0(\mathbf{k})$ needs to be self-consistently renormalized.

Matrices with RS structure such as $\mathbf{M}_{ab}=A\,\delta_{ab}+B\,(1-\delta_{ab})$ have inverses of the same RS structure:
\begin{eqnarray}
\mathbf{M}_{ab}^{-1}=\frac{A+(m-2)B}{(A-B)(A+(m-1)B)}\delta_{ab}-\frac{B}{(A-B)(A+(m-1)B)}(1-\delta_{ab}),\nonumber
\end{eqnarray}
where $m$ is the number of replicas. Using this form in equations (\ref{propag}) and (\ref{dressed}), in the limit $m\rightarrow1$, and since the Hartree solution is diagonal, the diagonal and off-diagonal parts of the above functions obey the following equations:
\begin{eqnarray}
\begin{array}{lcl}
\mathcal{G}(\mathbf{k})=\frac{\mathcal{G}_0(\mathbf{k})}{1+\mathcal{G}_0(\mathbf{k})\Sigma_{G}(\mathbf{k})},&\mbox{                  }&
\mathcal{F}(\mathbf{k})=-\frac{\Sigma_{\mathcal{F}}(\mathbf{k})\,\mathcal{G}^2(\mathbf{k})}{1-\mathcal{G}(\mathbf{k})\Sigma_{\mathcal{F}}(\mathbf{k})},\nonumber\\
\Sigma_{\mathcal{G}}(\mathbf{k})= \int \frac{d^2 p}{(2\pi)^{2}}\, \mathcal{D}_{G}\, (\mathbf{p}) \,\mathcal{G}(\mathbf{p}+\mathbf{k}),&\mbox{                  }&
\Sigma_{\mathcal{F}}(\mathbf{k})= \int \frac{d^2 p}{(2\pi)^{2}}\, \mathcal{D}_{F}\, (\mathbf{p}) \,\mathcal{F}(\mathbf{p}+\mathbf{k}),\nonumber\\
\mathcal{D}_{\mathcal{G}}(\mathbf{k})=\frac{v}{1+v\Pi_{\mathcal{G}}(\mathbf{k})},&\mbox{                  }&
\mathcal{D}_{\mathcal{F}}(\mathbf{k})=-\frac{\Pi_{\mathcal{F}}(\mathbf{k})\,\mathcal{D}_{\mathcal{G}}^2(\mathbf{k})}{1-\mathcal{D}_{\mathcal{G}}(\mathbf{k})\Pi_{\mathcal{F}}(\mathbf{k})},\nonumber\\
\Pi_{\mathcal{G}}(\mathbf{k})= \int \frac{d^2 p}{(2\pi)^2}\, \mathcal{G}(p)\, \mathcal{G}(\mathbf{p}+\mathbf{k}),&\mbox{                  }&
\Pi_{\mathcal{F}}(\mathbf{k})= \int \frac{d^2 p}{(2\pi)^2}\, \mathcal{F}(p)\, \mathcal{F}(\mathbf{p}+\mathbf{k}).
\end{array}
\end{eqnarray}